\documentclass[
twocolumn, preprintnumbers, amsmath,amssymb]{revtex4}

\usepackage{graphicx}
\usepackage{dcolumn}
\usepackage{bm}

\newcommand\ee{\end{equation}}
\newcommand\be{\begin{equation}}
\newcommand\eea{\end{eqnarray}}
\newcommand\bea{\begin{eqnarray}}
\newcommand{\sfrac}[2]{{\textstyle\frac{#1}{#2}}}
\newcommand\di{\partial}

\begin{document}


\title{Low-energy effective field theory for\\
finite-temperature relativistic superfluids}

\author{Alberto Nicolis}
\email{nicolis@phys.columbia.edu}

\affiliation{%
Physics Department and Institute for Strings, Cosmology, and Astroparticle Physics,\\
Columbia University, New York, NY 10027, USA
}%

\date{\today}

\begin{abstract}
We derive the low-energy effective action governing the infrared dynamics of relativistic superfluids at finite temperature. We organize our derivation in an effective field theory fashion---purely in terms of infrared degrees of freedom and symmetries. Our degrees of freedom are the superfluid phase $\psi$, and the comoving coordinates for the volume elements of the normal fluid component. The presence of two sound modes follows straightforwardly from Taylor-expanding the action at second order in small perturbations. 
We match our description to more conventional hydrodynamical ones, thus linking the functional form of our Lagrangian to the equation of state, which we assume as an input.
We re-derive in our language some standard properties of relativistic superfluids in the high-temperature and low-temperature limits. As an illustration of the efficiency of our methods, we compute the cross-section for a sound wave (of either type) scattering off a superfluid vortex at temperatures right beneath the critical one.

\end{abstract}

\maketitle

\section{Introduction}
The low-energy, long-distance dynamics of superfluids are usually studied at the level of hydrodynamical equations. At finite temperature, these describe two fluids interacting in a non-trivial way. For instance, they can exchange charge and mass on top of energy and momentum.  The standard derivation of the equations of motion and of the so-called constitutive relations is somewhat cumbersome, requires a clever use of galilean invariance, and an extensive one of thermodynamics (see e.g.~\cite{LL, putterman}). The generalization to relativistic superfluids is possible \cite{Israel, KL, Carter, CK, Israel2}, but not particularly transparent.

Here,  we present an alternative approach based on effective field theory. We extend  previous work on solids and ordinary fluids \cite{DGNR, ENRW, DHNS} to the case of (non-dissipative) relativistic superfluids. 
We see several advantages in our approach:

\begin{itemize}

\item 
The system is described by a field theory, in terms of a local action. The action encodes all the dynamics in the most compact way, as usual. The variational principle is the standard one for local field theories, with no Lagrange multipliers nor constraints nor additional restrictions on the fields' variations.

\item 
The starting point is the long-distance degrees of freedom and the symmetries acting on them. The action is the most general one compatible with the symmetries, organized as a derivative expansion. In particular, we will not try to cook up an ad-hoc action that {\em reproduces} the known hydrodynamical equations for superfluids. We will instead {\em derive} the same equations straightforwardly and systematically from a different set of principles---those of effective field theory.

\item
Implementing Poincar\'e invariance is straightforward: the local fields that parametrize the infrared degrees of freedom are Lorentz scalars. Laundau's clever use of galilean invariance is here replaced by the more prosaic ``contract the indices" rule. 

\item 
Thermodynamics is not needed. More precisely: Since the dynamics descend from an action, they are non-dissipative by construction. In such a case, thermodynamics is needed just to establish a dictionary between our field variables and the standard hydrodynamical and thermodynamical ones (sect.~\ref{thermo}). Once this is done, one can just forget about the thermodynamical interpretation of the system and use the action to compute observables.

\end{itemize}

Besides the physical importance of relativistic effects for certain superfluid systems---e.g.~for neutron stars---keeping the analysis fully relativistic from the start presents technical and conceptual advantages. At the technical level, it is just simpler and neater than the non-relativistic one---all our dynamical fields transform linearly in an irreducible representation of the Poincar\'e group (the scalar one), so that enforcing Poincar\'e invariance is trivial. By contrast, for instance, the velocity fields of the standard non-relativistic hydrodynamical description transform non-linearly under Galileo boosts. At the conceptual level, the relativistic analysis dodges certain somewhat misleading accidental degeneracies that arise in the non-relativistic limit but that are not enforced by the symmetries. For instance, in the non-relativistic limit mass density and mass current are proportional to charge density and charge current. When relativistic effects are included, the degeneracy is gone: `mass' is not conserved anymore---it is not even well defined---while charge is, as enforced by a $U(1)$ symmetry.
The non-relativistic limit can be taken straightforwardly when (and if) needed.

For genuinely hydrodynamical questions our approach is probably no better than the standard one. If, for instance, the goal is to solve the hydrodynamical equations with certain boundary conditions, then the fact that we have a neater way to get to those equations is of no use---we still have to solve them! There are however certain questions for which our formalism is manifestly more convenient. For instance, perturbation theory is straightforward and systematic, as usual for a local field theory with an action: In sect.~\ref{sounds} we show that there are two propagating sound modes. In sect.~\ref{scatter} we consider how these interact with a superfluid vortex. Computing the associated scattering cross-sections is straightforward, via the usual Feynman rules.

From the field-theoretical viewpoint, this paper builds on previous work on the field theory of
fluids \cite{DGNR, ENRW, DHNS} and of zero-temperature superfluids \cite{son1}, and carries similarities to ref.~\cite{supersolids}, where supersolids are discussed in effective field theory terms. From the hydrodynamical viewpoint, our analysis will mirror that of \cite{son2, HKS}, with a different starting point.


\section{The setup}
We begin by briefly reviewing the field theoretical description of a zero-temperature superfluid (see e.g.~\cite{son1}). In field-theory terms, a superfluid can be thought of as a system carrying a conserved $U(1)$ charge in a state that  {\em (i)} has finite density for this charge and that {\em (ii)} spontaneously breaks the corresponding $U(1)$ symmetry. The spontaneous breaking---or Bose-Einstein condensation---is in a sense physically {\em caused} by having finite charge density, but, once we consider a state that has both, we can discuss their implications in any order \cite{NPS}. We find it more convenient to start with the spontaneous breaking.
It implies the existence of a gapless excitation $\psi$---the Goldstone boson---which non-linearly realizes the $U(1)$ symmetry:
\be \label{shift}
\psi \to \psi + a \; . \qquad a = {\rm const}
\ee
Barring accidents, one can assume that the low-energy dynamics involve just this degree of freedom. The low-energy Lagrangian for $\psi$ should then be the most general one compatible with the shift-symmetry  \eqref{shift} as well with Poincar\'e invariance, organized as a derivative expansion. At lowest order in derivatives, it takes the form \cite{son1}
\be \label{PofX}
{\cal L} = P(X) \; , \qquad X \equiv \di_\mu \psi \di^\mu\psi \; ,
\ee
where $P$ is, for the moment, a generic function. The current associated with the $U(1)$ symmetry \eqref{shift} is 
\be
j^\mu = 2P'(X) \di^\mu \psi \; .
\ee
We see that for the superfluid state to have finite charge density, one needs finite $\dot \psi$. In a state of uniform charge density and vanishing spacial current,
\be \label{psi0}
\psi = \mu t \; ,
\ee
where $\mu$ is the chemical potential \cite{son1}. The function $P(X)$ turns out to be nothing but the equation of state, interpreted as a relation between the pressure $p$ and the chemical potential, $p = P(-\mu^2)$ \cite{son1}. Notice that from the field-theoretical viewpoint, the system's behaving as a fluid---that is, its obeying the equations of hydrodynamics---is quite non-trivial, and surprising. The corresponding fluid four-velocity field is $\di^\mu \psi$, suitably normalized. In particular, it obeys a relativistic generalization of the irrationality condition, i.e., it describes potential flow.

Now, Laundau's intuition is that a superfluid at {\em finite} temperature should be describable as a mixture of two mutually interacting fluids, one that behaves like the zero temperature superfluid we have just described, with irrotational flow only, and one that behaves like a normal fluid. The intuition is substantiated by the following considerations.
The homogeneous and isotropic state \eqref{psi0} admits gapless excitations $\pi$,
\be
\psi = \mu t + \pi \; ,
\ee
the superfluid phonons. At finite temperature, these will get excited, will reach thermodynamic equilibrium,  and will form a thermal bath. 
If one now considers out-of-equlibrium perturbations, with frequencies lower than the phonon inverse mean free time, and with wavelengths longer than their mean free path, this phonon bath will be describable by ordinary hydrodynamics. One thus has a normal fluid made up of phonons, moving in the presence of, and interacting with a background superfluid---which can also ``move'' of course. (As we stressed already, from the field theory viewpoint the hydrodynamical interpretation of the superfluid dynamics is somewhat accidental. By a moving superfluid we mean a field configuration different from \eqref{psi0}.) On the other hand, at much higher temperatures, above a critical temperature $T_c$, the spontaneous breaking will be gone, superfluidity will be lost,  and the dynamics  will be those of an ordinary fluid---like for any other system. 
It is thus natural to postulate that at any temperature between zero and $T_c$, the system be made up of two fluids---a `super' one and a `normal' one, and that the density of the former decreases monotonically with temperature, dropping to zero at $T_c$ and staying zero thereafter. This is, in essence, Landau's two-fluid model.

Taking for granted these physical facts, we seek a low-energy field theoretical description of such a two-fluid system.
We should first isolate the low-energy degrees of freedom and the symmetries acting on them. We should then construct the most generic Lagrangian involving the former and compatible with the latter, organized as a derivative expansion. The degree of freedom parameterizing the superfluid component is the scalar $\psi(\vec x, t)$ we have been considering so far. As we argued, it enjoys a shift simmetry---eq.~(\ref{shift}). The normal component has the usual long-wavelength degrees of freedom of ordinary hydrodynamics, which for field theoretical purposes are conveniently parameterized by three scalar fields $\phi^I(\vec x, t)$, with $I=1,2,3$ \cite{DGNR, ENRW}. These should be thought of as the comoving (or `Lagrangian') coordinates of the fluid element occupying physical (or `Eulerian') position $\vec x$ at time $t$. Their dynamics should be invariant under the following internal symmetries \cite{DGNR, ENRW}:
\begin{align}
\phi^I & \to \phi^I + a^I \; , & a^I = {\rm const} &  \label{shift2}\\ 
\phi^I & \to R^I {}_J \, \phi^J \; , & R \in SO(3) &\\
\phi^I & \to \xi^I (\phi^J ) \; , & \det \frac{\di \xi^I}{\di \phi^J} = 1 & \label{diff}\; .
\end{align}
The first two symmetries correspond to the physical homogeneity and isotropy of the fluid's internal space. The last symmetry is
what distinguishes a fluid from an isotropic solid---the dynamics' insensitivity to transverse deformations of the system.
The extra symmetry we should also impose is Poincar\'e invariance---which is straightforward to implement given that we are dealing with scalar fields. 

Given the shift-symmetries \eqref{shift} and \eqref{shift2}, our fields should enter the Lagrangian with at least one derivative acting on each of them. Therefore, at lowest order in the derivative expansion, the Lagrangian is made up of $\di_\mu \psi$ and $\di_\mu \phi^I$.
Moreover, in order to obey the volume-preserving diff symmetry \eqref{diff}, the fluid variables $\phi^I$ have to appear in the combination
\be \label{J}
J^\mu \equiv \sfrac16 \epsilon^{\mu\alpha\beta\gamma} \epsilon_{IJK} \di_\alpha \phi^I \di_\beta \phi^J \di_\gamma \phi^K \; .
\ee
With the two vectors $\di_\mu \psi$ and $J^\mu$, we can construct three scalar quantities---the three independent scalar products---compatible with all the symmetries. In fact, it turns out to be slightly more convenient to first factor out $J^\mu$'s norm:
\be \label{b}
b \equiv \sqrt{-J_\mu J^\mu} = \sqrt{ \det {\di_\mu \phi^I \di^\mu \phi^J}} 
\ee
($J^\mu$ is time-like---hence the minus sign.) The resulting normalized four-vector,
\be
u^\mu \equiv \frac1b J^\mu \; ,
\ee
is nothing but the fluid's four-velocity: it is normalized to $-1$, and, given eq.~(9), it is a vector field along which comoving coordinates do not change:
\be
u^\mu \di_\mu \phi^I = 0 \; , \qquad I=1,2,3 \; .
\ee
These properties {\em define} the fluid velocity field. The three invariant scalars we can construct at this order in derivatives therefore are
\be
b\; , \qquad X \equiv \di_\mu \psi \, \di^\mu \psi \; , \qquad y \equiv u^\mu \di_\mu \psi \; .
\ee
The low-energy effective Lagrangian should be a generic function of them:
\be
{\cal L} = F(b,X,y) \label{L} \; .
\ee
Our claim is that this field theory encodes the infrared dynamics of finite-temperature relativistic superfluids. 

Before checking this claim against more standard formulations of superfluid dynamics, we want to stress the economy of our approach, and the conceptual advantages that such economy offers. The degrees of freedom are parameterized by four scalar fields, with several internal symmetries. At low-energies, such symmetries are powerful enough to isolate three scalar combinations of our fields as the only possible invariants. The Lagrangian is a  function of these three quantities. Everything else follows  straightforwardly from the Lagrangian: the equations of motion (upon varying with respect to the fields), the stress-energy tensor (upon varying with respect to the metric), the conserved $U(1)$ current (via Noether's theorem), the spectrum of excitations (upon expanding at second order in small perturbations), the interactions of such excitations (via standard perturbation theory).

Moreover, we want to stress that unlike for previous Lagrangian approaches to fluid-dynamics, our Lagrangian is really a Lagrangian in the most ordinary sense. The equations of motion are obtained via the standard variational principle---we vary the fields while keeping their values at infinity fixed, without further restrictions on the variations. All fields have non-vanishing conjugate momenta, and, consequently, none of them plays the role of a Lagrange multiplier. This is achieved by treating the positions of the fluid elements as our dynamical variables, rather than bypassing them and trying to construct an action directly for apparently more natural degrees of freedom like the velocity and the density fields. In our approach these are derived quantities, precisely like in Lagrangian mechanics the velocity $\dot q(t)$ is a derived quantity, while it is the coordinate $q(t)$ that  plays the role of the fundamental `field' entering the action principle.


\section{First and second sound} \label{sounds}
As we will see, because of the unconventional language that we are using, the most laborious part of our analysis will be matching our results to the standard ones. However, to convince the reader of  the usefulness of our approach, we can show straightforwardly that the spectrum of excitations is composed of two propagating longitudinal sound modes, as befits a superfluid at finite temperature.

Consider an isotropic and homogeneous background configuration, where $\psi$ takes the form \eqref{psi0}, and the normal fluid part is at rest and homogeneous in such a reference frame. This means that the comoving coordinates $\phi^I$ are aligned with the physical ones, $\phi^I \propto x^I$. The proportionality constant  measures the compression level of such an homogeneous state, and depends, for instance, on the applied external pressure. Now consider small perturbations about such a background:
\be
\psi(x) = y_0 \cdot (t + \pi^0 (x)) \; , \qquad \phi^I (x) = b_0^{1/3} \cdot (x^I + \pi^I (x)) \; .
\ee
The proportionality constants have been chosen for notational convenience, since
\be
y = y_0 + {\cal O}(\pi) \; , \qquad b = b_0 + {\cal O}(\pi) \; .
\ee
To study the free propagation of small perturbations, one has to expand the Lagrangian at quadratic order in $\pi^0$ and $\vec \pi$. After a straightforward Taylor expansion, we get 
\begin{align} \label{L2}
{\cal L} & \simeq \sfrac12 \big[ K_N \dot{\vec \pi}^2 - G_N (\vec \nabla \cdot \vec \pi)^2 \big] \nonumber \\
   &+ \sfrac12 \big[ K_S( \dot \pi^0)^2 - G_S (\vec \nabla \pi^0)^2   \big] \nonumber  \\
   & + M \, (\vec \nabla \cdot \vec \pi) \dot \pi^0 
\end{align}
where the various coefficients are defined as
\begin{align}
K_N & = (F_y y_0 -F_b b_0) \label{K_N} \\
G_N & = -F_{bb} b_0^2 \\
K_S & = (F_{yy}-2 F_X) y_0^2 -4 F_{yX}y_0^3 +4F_{XX}y_0^4 \\
G_S & = -2 F_X y_0^2 \\
M & = F_{by}b_0y_0-F_y y_0 -2 F_{bX} b_0 y_0^2 \label{M}
\end{align}
On the right-hand-sides, the subscript $b$, $y$, and $X$ stand for partial derivatives of $F$---all computed at the background values $b=b_0$, $y=y_0$, and $X=-y_0^2$. On the left-hand-sides,
$K$, $G$, $N$, $S$, and $M$  stand for `kinetic', `gradient', `normal', `super', and `mixing', respectively.

Despite the apparent complexity of these coefficients, the structure of the quadratic Lagrangian \eqref{L2} is extremely simple. We can see right away that, precisely like for an ordinary fluid, the transverse (i.e., divergence-less) component of $\vec \pi$ does not propagate. It enters the quadratic Lagrangian as
\be
{\cal L} \supset \sfrac12 K_N \dot{\vec \pi}_T^2 \; ,
\ee
that is, there is no gradient energy for it, nor does it get one from mixing with $\pi^0$---this is a direct consequence of the volume-preserving diff symmetry \eqref{diff}. As a result, $\vec \pi_T$ does not feature propagating wave solutions. (For a more in-depth discussion about this point, we refer the reader to \cite{ENRW}.)

On the other hand,  $\vec \pi$'s longitudinal component and $\pi^0$ have standard kinetic and gradient energies. Moreover, there is a term mixing them---that weighed by $M$. Given that $F$ is a generic function---we will see that it is related to the equation of state, which can be fairly generic---we should think of all the coefficients \eqref{K_N}--\eqref{M} as independent, since they involve different combinations of derivatives of $F$. As a result, barring accidental cancellations, we expect that the diagonalization of the $\vec \pi_L$-$\pi^0$ system yield two non-degenerate propagating modes, with eigenfrequencies determined by the secular equation
\be \label{secular}
\det \left( \begin{array}{cc} 
K_N \omega^2 -G_N k^2  &  M \, \omega k \\
M \, \omega k& K_S \omega^2 -G_S k^2 
\end{array}
\right) = 0 \; .
\ee

We thus have two independent kinds of sound waves, propagating non-dispersively (i.e., with $\omega \propto k$) at different speeds, and corresponding to different linear combinations of a longitudinal deformation $\vec \pi_L$ of the normal fluid component, and of a perturbation $\pi^0$ of the superfluid one. These are precisely the qualitative properties of first and second sound. We will see that our description works quantitively as well---for instance the propagation speeds match precisely those computed via Landau's hydrodynamic theory---but before being able to do so, we have to find the dictionary connecting our field-theoretical language to hydrodynamic and thermodynamic variables.

%


%


\section{Hydrodynamics and thermodynamics}\label{thermo}
We will follow the same logic as in \cite{DHNS}.
We start by computing the stress-energy tensor associated with the Lagrangian \eqref{L}. Let's consider an infinitesimal variation of the spacetime metric, $g^{\mu\nu} = \eta^{\mu\nu} + \delta g^{\mu\nu}$. We should compute the associated variation of the action, when we keep all our field variables $\psi, \phi^I$ constant. The metric enters the action in several ways. First of all, there is the overall $\sqrt{-g}$ upfront. Second, $b$, $X$, and $y$ all depend on the metric. For $b$, from the r.h.s~of eq.~\eqref{b} we have
\be
\delta b  = \sfrac12  b \,  B^{-1}_{IJ} \, \di_\mu \phi^I \di_\nu \phi^J \, \delta g^{\mu\nu} \; ,
\ee
where $B^{-1}_{IJ}$ is the inverse of 
\be
B^{IJ} \equiv \di_\mu\phi^I \di^\mu \phi^J \;  ,
\ee
and we made use of the identity
\be
b^2 = \det B^{IJ} \; .
\ee
For $X$, we have simply
\be
\delta X = \di_\mu \psi \di_\nu \psi \, \delta g^{\mu\nu} \; .
\ee
For $y$, the metric enters through $u^\mu = J^\mu/b$, which depends on it both because of $b$, and because of the $\epsilon$-symbol in $J^\mu$, $\epsilon^{\mu\alpha\beta\gamma} \propto \frac{1}{\sqrt{-g}}$. Collecting all these contributions we get
\begin{align}
T_{\mu\nu} &  = - 2\frac{\delta S}{\delta g^{\mu\nu}} \nonumber \\
& = \big(F_y y - F_b b \big) B^{-1}_{IJ} \di_\mu \phi^I \di_\nu \phi^J + (F-F_y y) \eta_{\mu\nu} \nonumber \\
&\quad - 2 F_X \di_\mu \psi \di _\nu \psi  \; .
\end{align}
Using the identity
\be
B^{-1}_{IJ} \di_\mu \phi^I \di_\nu \phi^J = \eta_{\mu\nu} + u_\mu u_\nu \; ,
\ee
we can rewrite this as
\be \label{Tmn}
T_{\mu\nu}  = \big(F_y y - F_b b \big) u_\mu u_\nu + (F- F_b b) \eta_{\mu\nu} - 2 F_X \di_\mu \psi \di _\nu \psi  \; .
\ee

Next, we compute the current associated with the spontaneously broken $U(1)$ symmetry \eqref{shift}. The Lagrangian depends on $\psi$ only through $\di_\mu \psi$. In such a simple case, the shift Noether current reduces to
\begin{align} 
j^\mu & = \frac{\di {\cal L}}{\di (\di_\mu \psi)} \nonumber \\
 & = F_y u^\mu + 2 F_X \, \di^\mu \psi \; .   \label{j}
\end{align}

Now, to match our field-theory quantities to the more standard hydrodynamical and thermodynamical ones, we have to extract the energy density $\rho$, the pressure $p$, and the particle number density $n$, from $T_{\mu\nu}$ and $j^\mu$ above, and then to impose the usual thermodynamics identities. To extract $\rho$ and $n$ we have to pick a local reference frame. Since we have two fluids in relative motion, there are {\em two} natural reference frames, which means that neither is really {\em the} preferred one. Ultimately it does not matter which frame we pick as long as we apply thermodynamics consistently in that frame. We choose to move with the normal fluid component, along $u^\mu$, also because---as already emphasized---the interpretation of $\di^\mu \psi$ as a superfluid velocity field is perhaps possible, but not necessary. We have
\begin{align}
\rho & \equiv T^{\mu\nu} \, u_\mu u_\nu = F_y y - F -2 F_X y^2 \\
n & \equiv - j^\mu u_\mu = F_y - 2 F_X y \label{n}
\end{align}
(the minus sign in the definition of $n$ takes care of the negative normalization of $u^\mu$.) As to the pressure, unlike for a single fluid, there is in general no frame in which the stress-tensor is isotropic. This makes the definition of pressure less obvious than usually. However, $T_{\mu\nu}$ is still the sum of an isotropic tensor ($\propto \eta_{\mu\nu}$)
and of rank-one tensors, one proportional to $u_\mu u_\nu$, and another one proportional to $\di_\mu \psi \di _\nu \psi$.
This decomposition is unambiguous, and invites identifying $p$ with the coefficient of $\eta_{\mu\nu}$. This means that in any reference frame, and for non relativistic fluid motions, the stress tensor $T_{ij}$ will be the sum of $p \,\delta_{ij}$ and of two anisotropic terms, proportional to $v^i_N v^j_N$ and to $v^i_S v^j_S$, where $\vec v_N$ and $\vec v_S$ are the normal fluid and superfluid velocity fields. This is the standard definition of pressure for non-relativistic superfluids, and from the thermodynamical viewpoint it corresponds to the derivative of the total energy w.r.t.~the volume, taken at constant total entropy, charge, and relative momentum \cite{putterman}.
We are thus led to set
\be \label{p}
p = F - F_b b \; .
\ee

We now impose the thermodynamic identities. We start with 
\be
\rho + p = Ts + \mu n \; . 
\ee
Plugging in the expressions for $\rho$, $p$, and $n$ derived above we get
\be
\big( F_y -2 F_X y \big ) y - F_b b  = Ts + \mu \big( F_y - 2 F_X y \big) 
\ee
This suggests the identifications
\be \label{identify}
\mu = y \; , \qquad s = b\; , \qquad T = - F_b \; ,
\ee
which are consistent with Landau's idea that the entropy is carried by the normal fluid component only: $b$ does not depend on the superfluid field variable $\psi$, but only on the normal fluid ones, $\phi^I$. 
Moreover, as emphasized in \cite{DHNS}, with these definitions the entropy current coincides with our $J^\mu = b u^\mu$, which is {\em identically} conserved by virtue of its definition \eqref{J}.
There is course an ambiguity in the overall normalization of $s$ and $T$---we can multiply $s$ by a constant and divide $T$ by the same constant without affecting the thermodynamic identities. This just amounts to changing $k_B$, that is, to a change of units for temperature.

Then, we should impose one of the differential thermodynamic identities---for instance
\be
dp = s \, dT + n \, d \mu +  \dots
\ee
The dots stand for an additional term that takes into account that here the number of independent thermodynamic variables is {\em three}, rather than the usual two. This is evident from our field-theoretical description: the dynamics encoded by the Lagrangian \eqref{L} involve three independent scalar quantities. We have already identified $b$ with the entropy density, and $y$ with the chemical potential. We are still missing the thermodynamic interpretation of $X$. In more standard treatments of (non-relativistic) superfluid dynamics, the extra independent variable is the relative velocity---or the relative momentum density---between the superfluid and normal fluid parts. We will recover this interpretation below.

If we compute the differential of \eqref{p} in terms of $dT = - dF_b$, $d\mu = d y$, and $dX$, we get
\begin{align}
dp &= - b \, dF_b + F_y \, dy + F_X \, dX \nonumber \\
 & = s \, dT + n \, d \mu  + F_X (dX + 2 y \, dy) \; .
\end{align}
The combination in parentheses is most easily written in terms of the components of $\di_\mu \psi$ orthogonal to $u^\mu$:
\be
\di_\mu \psi = - u_\mu \, y + \xi_\mu \; , \qquad \xi^\mu \equiv (\eta^{\mu\nu}+u^\mu u^\nu) \di_\nu \psi \; .
\ee
$\xi^\mu$, being orthogonal to $u^\mu$, can be thought of as a three-vector $\vec \xi$.
We have $X = - y^2 + \xi^2$, so that
\be \label{dp}
dp = s \, dT + n \, d \mu  + 2 F_X \xi  \, d \xi \; ,
\ee
with $\xi \equiv |\vec \xi \, |$.

At the level of hydrodynamics and thermodynamics, our description of the system is manifestly equivalent to that of Son et al.~\cite{son2, HKS}, which in turn is equivalent to those of Israel \cite{Israel}, of Khalatnikov and Lebedev \cite{KL}, and of Carter \cite{Carter} (see \cite{CK, Israel2} for the equivalence of these different classic approaches). To see this, apart from obvious changes in the notation, one only needs to take into account that $n$ and $\rho$ here stand for the total number and energy densities. On the other hand, in \cite{son2, HKS} $n$ stands for the {\em normal} component's number density---our $F_y$ in eq.~\eqref{n}---while $\epsilon$ is {\em defined} via $\epsilon+p = T s + \mu n$ \footnote{Notice that it is not clear what $\epsilon$ thus defined corresponds to, physically. For instance, at zero temperature entropy vanishes and so does the normal component's charge density, so that $\epsilon $ becomes negative: $\epsilon = - p$.}.

To summarize, we see from eq.~(\ref{dp}) that pressure is naturally a function of $T$, $\mu$, and $\xi$. Such a function defines the equation of state of our superfluid. Once it is given, one can construct the corresponding Lagrangian (\ref{L}) from eq.~\eqref{p}, that is
\begin{align}
{\cal L} = F &= p + F_b b \\
             & =  p - T \frac{\di p}{\di T}   \; ,
\end{align}
where we used eqs.~\eqref{identify} and \eqref{dp}. The result should then be expressed in terms of $b=s=\frac{\di p}{\di T}$, $y=\mu$, and $X = \xi^2-y^2$. We will see an explicit example of this procedure below.




\subsection{Relative motion}
One of the surprising features of superfluids at finite temperature is the possibility of having, {\em in thermal equilibrium}, a relative motion between the two components. The two fluids can freely flow through each, provided their motion is homogeneous and  happens at constant speed.
In our language, a necessary condition for this to be possible is that our field equations admit stationary and homogeneous solutions where the reference frame defined by the superfluid---which is associated with the vector field $\di_\mu \psi$---is at not rest with respect to the normal fluid's one---which is associated with the vector that we call $u^\mu$. 
In other words, there should be solutions of the form,
\be \label{relative}
\phi^I (\vec x ,t) = b_0 \, x^I \; , \qquad  \psi (\vec x, t)= y_0 \, t + \vec \xi_0 \cdot \vec x \; ,
\ee
where $b_0$, $y_0$, and $\vec \xi_0$ are constant. In particular, $\vec \xi_0$ is the same variable we introduced in the last section, and it quantifies the misalignment between the two reference frames:
\be
u^\mu = (1, \vec 0) \; , \qquad \di_\mu \psi = (y_0, \vec \xi_0) \; .
\ee

It is easy to see that {\em all} configurations of the form \eqref{relative} solve our field equations.
The reason is that the Lagrangian \eqref{L} only depends on first derivatives of the fields. As a result, the field equations all take the schematic form
\be
\di \,G( \di \phi^I, \di \psi) = 0 \; .
\ee
Such equations are obviously obeyed by the field configuration \eqref{relative}, which has constant first derivatives.

The existence of such solutions is only a necessary condition for relative motion's being compatible with thermal equilibrium. This is because our field theory neglects dissipative effects, which are clearly crucial for reaching equlibrium.
What our solutions show is the absence of conservative forces between the two fluids in relative motion.
However as we argued above, for our field theory to describe a thermal system, this has to be characterized by three thermodynamic variables, one of which is precisely $\vec \xi$. In other words, the equation of state---which is a statement about the system {\em at equilibrium}---depends  on the state of relative motion. This means that relative motion is compatible with thermal equilibrium.

\section{The high-temperature limit}
The high temperature behavior of a superfluid is characterized by the disappearance of superfluidity at some critical temperature $T_c$. Right beneath $T_c$, the superfluid component is extremely small, and we can work perturbatively in it. That the superfluid component is small, concretely means that its contributions to the stress-energy tensor and to the charge current are small. From eqs.~\eqref{Tmn}, \eqref{j} we see that this corresponds to having small $F_X$: in the limit in which $F_X$ vanishes, the stress-energy tensor and the current reduce to those of an ordinary fluid flowing along $u^\mu$, which for us is the normal component's four-velocity field. We are thus led to the conclusion that the dynamics of a superfluid close to the transition temperature are captured by the behavior of our field theory  in a region of field space where $F_X$ is small. In fact, at $T=T_c$ and above, the superfluid component disappears completely, and one is left with an ordinary fluid carrying a conserved charge. In our Lagrangian formalism, this is described by \cite{DHNS}
\be \label{Fofby}
{\cal L}_{T>T_c} = F(b,y) \; .
\ee
This means that above $T_c$ there is no dependence on $X$ whatsoever. As pointed out in \cite{DHNS}, this is equivalent to imposing the generalized shift-symmetry
\be
\psi \to \psi + f(\phi^I)
\ee
for generic $f$, under which $b$ and $y$ are invariant, but $X$ is not. Therefore, even though it looks ``unnatural'' to have a whole region of $(b,y,X)$-space where $F$ does not depend on $X$, such a region is in fact one of enhanced symmetry. As a result, its existence is protected by symmetry.

Let's consider the case described by \eqref{Fofby} first. For small perturbations about the equilibrium configuration, the quadratic Lagrangian \eqref{L2} simplifies to
\begin{align}
{\cal L} & \simeq  \sfrac12 \big[ (F_y y_0 - F_b b_0) \dot {\vec \pi}^2 + F_{bb} b_0^2 \, (\vec \nabla \cdot \vec \pi)^2 \big] \nonumber \\
& + \sfrac12 F_{yy} y_0^2 \, (\dot \pi^0)^2 + (F_{by} b_0 y_0 - F_y y_0) (\vec \nabla \cdot \vec \pi ) \dot \pi^0 \label{L2aboveTc} \; .
\end{align}
Notice that now there is no gradient energy for $\pi^0$, but there is still a mixing between $\pi^0$ and $\vec \pi_L$. The Lagrangian is easily diagonalized via the field redefinition
\begin{align} \label{pi1pi2}
\dot \pi^0 & = \dot \pi_2 -\sfrac{F_{by} b_0 - F_y}{F_{yy} y_0} (\vec \nabla \cdot \vec \pi_1) \; , \\
\vec \pi_L & = \vec \pi_1 \; ,
\end{align}
where $\vec \pi_1$ is purely longitudinal. Neglecting $\vec \pi_T$---which does not propagate---we get
\begin{align}
{\cal L} & \to \sfrac12 \big[ (F_y y_0 - F_b b_0) \dot {\vec \pi}_1^2 + \sfrac{ F_{yy} F_{bb} b_0^2 - (F_{by} b_0 - F_y)^2}{F_{yy} } \, (\vec \nabla \cdot \vec \pi_1)^2 \big]  \nonumber \\
& + \sfrac12 F_{yy} y_0^2 \, \dot \pi_2^2 \; . 
\end{align}
We see that $\pi_2$---like $\vec \pi_T$---does not have a gradient energy term. As a consequence, it does not feature wave solutions. On the other hand, $\vec \pi_1$ has wave solutions propagating at (squared) speed
\be \label{cs}
c^2_1 = \frac{(F_{by} b_0 - F_y)^2- F_{yy} F_{bb} b_0^2 }{F_{yy} (F_y y_0 - F_b b_0) } \; .
\ee
Equivalently, when all derivatives of $F$ with respect to $X$ vanish identically, the secular equation \eqref{secular} has a non-trivial solution, $\omega^2_1 = c_1^2 k^2$, corresponding to waves of  $\vec \pi_1$, 
and a trivial one, $\omega^2_2 = 0$, corresponding to $\pi_2$'s trivial dynamics.

We thus see that above $T_c$ there is only one propagating sound mode, as befits an ordinary fluid. It is somewhat non-trivial to check that the propagation speed we found---eq.~\eqref{cs}---matches the standard hydrodynamic expression
\be \label{cs_hydro}
c_s^2 = \frac{\di p}{\di \rho}\bigg|_{S,N} \; ,
\ee
where $S$ and $N$ are the total entropy and charge. The reason is that our Lagrangian is naturally a function of $b=s$ and $y=\mu$, rather than of $s$ and $n$---and so are $\rho$ and $p$ in our formalism. A similar complication was encountered in \cite{HKS}, and  it was overcome by using thermodynamic identities. Here, we can use the following trick. In \eqref{cs_hydro}, the constraint of being at constant $S$ and at constant $N$ gives a one-to-one relation between $dp $ and $d \rho $. In our field theory, the same relation can be obtained via the equations of motion: the dynamics of our system conserve the total entropy by construction (since they follow from a conservative Lagrangian), and the total charge because of a $U(1)$ symmetry. Whatever relation we get between $dp$ and $d \rho$ by enforcing the equations of motion, is thus going to be equivalent to that enforced in \eqref{cs_hydro} \footnote{To be precise, the relation we get from the eom has to be {\em compatible} with that of  \eqref{cs_hydro}. However, since in \eqref{cs_hydro} we already have a unique relation between $dp$ and $d \rho$, if from the eom we are also able to get a unique $dp$-$d\rho$ relation, then the two relations have to coincide.}. 
Charge current  conservation reads
\be
\di_\mu (F_y u^\mu) = 0 \; . 
\ee
(we are setting $F_X=0$). By using $u^\mu = b J^\mu$, and the conservation identity for $J^\mu$ (which is equivalent to the conservation of the entropy current), we get
\be
J^\mu \di_\mu \big( F_y/ b \big) = 0
\ee 
That is, $( F_y/ b)$ is conserved along the flow, or,
\be
b \, dF_y = F_y \,db \qquad \mbox{(along the flow)} \; .
\ee
We thus see that the equations of motion imply a relation between $dF_y$ and $db$. This in turn translates into a relation between  $dp$ and $d\rho$. To see this, let's first express $dF_y$ in terms of $dy$ and $db$:
\be
dF_y = F_{yy}dy + F_{yb} db   \; .
\ee
This implies a relation between $dy$ and $db$ along the flow;
\be
b F_{yy} \, dy =(F_y - F_{yb} b) db  \qquad \mbox{(along the flow)} \; .
\ee
We thus get
\begin{align}
\frac{dp}{d\rho}\bigg|_{S,N} & = \frac{dp}{d\rho}\bigg|_{\rm flow} \nonumber \\
& = \frac{d(F - F_b b)}{d(F _y y -F)}\bigg|_{\rm flow} \nonumber \\
& =\frac{(F_{by} b - F_y)^2- F_{yy} F_{bb} b^2 }{F_{yy} (F_y y - F_b b) } \; ,
\end{align} 
which is precisely eq.~\eqref{cs}.

At temperatures slightly below $T_c$, $F$ acquires a weak $X$-dependence. Derivatives of $F$ with respect to $X$ are now non-vanishing, but can be treated as small. In particular, in the coefficients entering the quadratic action for small excitations, eqs.~\eqref{K_N}--\eqref{M}, we can ignore these $X$-derivatives everywhere except for $G_S$. The reason is that $G_S$ vanishes when these derivatives do, and as a consequence the secular equation \eqref{secular} becomes degenerate---i.e., one of the solutions becomes trivial, and one mode stops propagating. When $G_S \sim F_X$ is small but non-vanishing, this degenerate mode will start propagating, with a frequency $\omega_2^2 = {\cal O}(F_X) k^2$. Keeping into account the $X$-derivatives of $F$ for the other coefficients would simply shift the two eigenfrequencies by relatively negligible  amounts. That is, schematically:
\begin{align}
\omega_1^2 & = \big(c_1^2 + {\cal O}(F_X) \big) k^2 \\
\omega^2_2 & =  {\cal O}(F_X) \big( 1+ {\cal O}(F_X) \big)  k^2 \; .
\end{align}
To compute the speed of the new propagating mode at leading order, we just have to keep $F_X$ in $G_S$ and drop all other $X$-derivatives from the other coefficients. The relevant quadratic Lagrangian is then eq.~(\ref{L2aboveTc}) supplemented with a small gradient energy for $\pi^0$:
\begin{align}
{\cal L} & \simeq  \sfrac12 \big[ (F_y y_0 - F_b b_0) \dot {\vec \pi}^2 + F_{bb} b_0^2 \, (\vec \nabla \cdot \vec \pi)^2 \big] \nonumber \\
& + \sfrac12 \big[ F_{yy} y_0^2 \, (\dot \pi^0)^2 + 2 F_X y_0^2 \, (\vec \nabla \pi^0)^2 \big] \nonumber \\
& + (F_{by} b_0 y_0 - F_y y_0) (\vec \nabla \cdot \vec \pi ) \dot \pi^0 \label{L2belowTc} \; .
\end{align}
The near vanishing eigenfrequency of the secular equation becomes
\begin{align}
\omega^2_2 & \simeq -\frac{G_N G_S}{ M^2 + G_N K_S} k^2  \nonumber \\
& \simeq \frac{2F_{bb}  \, b_0^2 y_0^2}{(F_{by}  b_0y_0 - F_y y_0)^2 - F_{bb} F_{yy} \, b_0^2 y_0^2} \, F_{X} \,  k^2
\end{align}
It is straightforward but somewhat tedious to check that in the non-relativistic limit, this reduces precisely to the second sound dispersion law as predicted in Landau's hydrodynamic theory for superfluids near $T_c$ \cite{LL}. 

For the same reasons as above, for finite but small $F_X$ the eigenmodes are still $\vec \pi_1$ and $\pi_2$ as defined in \eqref{pi1pi2}, plus small ${\cal O}(F_X)$ corrections which we can neglect at leading order.


\section{The low-temperature limit}
Things are much neater at very low temperatures. At exactly zero-temperature, we just have the superfluid component, and the Lagrangian depends on $X$ only, as in eq.~\eqref{PofX}:
\be
{\cal L}_{T=0} = \bar F(X) \; .
\ee
All kinetic coefficients \eqref{K_N}--\eqref{M} vanish except for
\be
K_S = 4 \bar F_{XX} y_0^4 - 2 \bar F_X y_0^2\;, \qquad G_S = -2 \bar F_X y_0^2 \; .\
\ee
The propagation speed of $\pi^0$ waves is thus
\be \label{c1}
c^2_1 = \frac{\bar F {}_X - 2 \bar F_{XX} y_0^2}{\bar F_X} = \frac{dp}{d \rho} \;, 
\ee
where the subscript `1' stands for `first sound'. Notice that $dp/d \rho$ here is unambiguous, since both $p$ and $\rho$ depend on $X$ only.

At very small but non-vanishing temperatures, these phonons get excited a make up a normal fluid. This changes the equation of state, and consequently the Lagrangian gets slightly modified
\be
{\cal L} = \bar F(X) + \delta F(b,y,X)\; .
\ee
The entropy density is dominated by the phonon contribution, which scales as $s \sim T^3$. Since for us $s=b$ and $T= - \delta F_b$, we expect that at low temperatures
\be \label{b43}
\delta F = b^{4/3} f(y,X) \qquad b \to 0 \; .
\ee
We will see below that this is indeed the case.
All kinetic coefficients \eqref{K_N}--\eqref{M} get modified by ${\cal O}(b^{4/3}) = {\cal O}(T^4)$ corrections. However, for the same reasons as those analyzed above, we need to keep track of these corrections only for those coefficients that would vanish otherwise, which are $K_N$, $G_N$, $M$. Moreover, since $M$ contributes an off-diagonal term to the kinetic matrix, it appears quadratically in the secular equation, \eqref{secular}, while $K_N$ and $G_N$ appear linearly.
$M$ can thus be neglected at leading order.
The low-temperature quadratic Lagrangian then is
\begin{align}
{\cal L} & \simeq  \sfrac12 \big[ (
\delta F_y y_0 - \delta F_b b_0) \dot {\vec \pi}^2 + \delta F_{bb} b_0^2 \, (\vec \nabla \cdot \vec \pi)^2 \big] \nonumber \\
& + \sfrac12 \big[ (4 \bar F_{XX} y_0^4-2 \bar F_X y_0^2) (\dot \pi^0)^2 + 2 \bar F_X y_0^2 (\vec \nabla  \pi^0 )^2 \big]  \nonumber \; .
\end{align}
It is already in a diagonal form---the two sound modes coincide with $\pi^0$ and $\vec \pi_L$, and have propagating speeds $c_1^2$ as given in \eqref{c1} and
\be \label{c2}
c_2^2 = - \frac{\delta F_{bb} \, b_0^2}{\delta F_y y_0 - \delta F_b b_0} \; ,
\ee
respectively. 

Notice that in this limit the second sound corresponds to longitudinal perturbations of the phonon gas. Even though such a gas fades away as $T$ approaches zero, the second sound  speed approaches a finite value. Indeed, for non-relativistic superfluids Landau predicted that it should approach the limiting value $c_2^2 \to c_1^2/3$. This result has been generalized to relativistic superfluids in \cite{CL}, where the same limiting value was found.
To see whether our expression for $c_2^2$ reproduces this result, we need the low-temperature equation of state. Fortunately the finite-temperature part of such an equation of state---which is associated with our $\delta F$---is model-independent. The reason is that it is determined purely by the low-temperature behavior of the phonon gas. Since phonons are derivatively coupled to each-other, at low temperatures they behave as free particles propagating at speed $c_1$. $c_1$ is thus the only parameter of the underlying zero-temperature superfluid entering the leading order finite temperature correction to the equation of state.

The low-temperature equation of state for relativistic superfluids has been computed in \cite{CL}. In our notation, it relates the pressure to $b$, $y$, and $X$:
\be
p \simeq \bar F (X)+ \bigg[\frac{b^4}{c_1} \Big(1+ (1 - c_1^2)\frac{y^2}{X} \Big)^{2} \bigg]^{1/3} \; .
\ee
Using \eqref{p} we get
\be
\delta F(b,y,X) \simeq -3 \bigg[\frac{b^4}{c_1} \Big(1+ (1 - c_1^2)\frac{y^2}{X} \Big)^{2} \bigg]^{1/3} \; ,
\ee
which, indeed,  is of the form \eqref{b43}.
Evaluating the derivatives entering \eqref{c2} at the background values $b_0$, $y_0$, $X_0=-y_0^2$, we get
\be
c_2^2 = \sfrac13 c_1^2 \; ,
\ee
as expected.

\section{Sound wave scattering off a superfluid vortex} \label{scatter}

So far we have been preoccupied with trying to dignify our field-theoretical approach by making it reproduce well known results about superfluids. Here instead, we make it compute something new. One of the main advantages of a field-theoretical Lagrangian description, is that it offers a systematic and straightforward approach to perturbation  theory, e.g.~for the study of scattering processes. With our superfluid theory we can consider  the scattering of  sound waves colliding with each other, along the lines of \cite{ENRW}. Perhaps more interesting is the scattering process involving a sound wave and a superfluid vortex, which has been claimed to be a powerful probe of `quantum vorticity' \cite{LS}. A similar process has been considered in \cite{supersolids} for supersolids.

Consider a superfluid vortex configuration in an otherwise homogeneous, still superfluid:
\begin{align} 
\psi (\vec x, t)& = y_0 t + \psi_V \; , \qquad \psi_V(\vec x, t) = C \, \varphi \label{vortex1} \\ 
\phi^I (\vec x, t) & = b_0^{1/3} x^I \; . \label{vortex2}
\end{align}
We are assuming that the vortex is aligned with the $z$-axis, and we are denoting the azimuthal angle by $\varphi$. $C$ is a constant that determines the periodicity of $\psi$: since $\psi$ is the Goldstone boson for a spontaneously broken $U(1)$ symmetry, it should be periodic, $\psi \sim \psi + 2 \pi C$. Without loss of generality, we can set
\be
C = 1 \; .
\ee
This just fixes the normalization of $\psi$, and of the associated charge.

Before considering the scattering of sound waves off the vortex, we have to make sure that the vortex configuration is a solution. Since the presence of the vortex perturbs the spatial derivatives of $\psi$ only, $y$ and $b$ are unaffected by it:
\be
y = u^\mu \di_\mu \psi = y_0 \;, \qquad b= b_0 \; .
\ee
On the other hand, $X$ is perturbed substantially:
\be
X = (\di \psi)^2= - y_0^2 + \frac{1}{r^2} \; .
\ee
This suggests that if the Lagrangian did not depend on $X$, then (\ref{vortex1}, \ref{vortex2}) would be a solution. To check that this is indeed correct, consider the Lagrangian \eqref{Fofby}. As we mentioned, it enjoys a generalized shift-symmetry, $\psi \to \psi + f(\phi^I)$.
Now, the vortex configuration (\ref{vortex1}, \ref{vortex2}) can be obtained by applying one such transformation to the background configuration
\be \label{background}
\psi_0(\vec x,t) =  y_0 t \; , \qquad \phi_0^I (\vec x, t) = b_0^{1/3} x^I \; ,
\ee
with
\be
f(\phi^I) =  \arctan \phi_0^2/\phi_0^1 = \varphi.
\ee
Therefore, since the background \eqref{background} solves the equations of motion, so does the vortex configuration.

Now, the Lagrangian \eqref{Fofby} is the correct description of the system at or above the critical temperature. At temperatures right below it, the $X$-dependence of $F$ can still be treated as small, which means that the true vortex solution will be slightly different from (\ref{vortex1}, \ref{vortex2}). At leading order we can neglect this small difference and stick with (\ref{vortex1}, \ref{vortex2}). At significantly lower temperatures, there will still be vortex solutions, but they will differ significantly from (\ref{vortex1}, \ref{vortex2}). In particular, we expect the normal component---parametrized by the $\phi^I$'s---to be significantly perturbed. The study of these more complicated vortices is beyond the scope of this paper. We focus on  temperatures right beneath the critical one, where (\ref{vortex1}, \ref{vortex2}) is the leading order vortex solution.

To study the scattering of a sound wave off the vortex, we need the quadratic Lagrangian for small perturbations in the presence of the vortex:
\be
\psi = y_0 (t +  \pi^0) +  \psi_V \; , \qquad \phi^I  = b_0^{1/3} \cdot (x^I + \pi^I) \; .
\ee
At leading order in the $X$-derivatives of $F$, we get \eqref{L2belowTc} for the vortex-independent part of the quadratic Lagrangian, 
and, after some integration by parts,
\begin{align}
{\cal L}_{\rm int} & \simeq \big[(F_y-F_{yb} b_0) (\vec \nabla \cdot \vec \pi) - F_{yy} y_0 \, \dot \pi^0 \big] \dot{\vec \pi} \cdot \vec \nabla \psi_V \nonumber \\
& -\sfrac12 F_y \, \dot \pi^i \pi^k \, (\di_i \di_k -\di_k \di_i) \psi_V \; 
\end{align}
for the vortex-sound wave interaction part. The Taylor expansion leading to this expression is straightforward. We have omitted 
${\cal O}(\di_i \psi_V)^2$ terms, since their relative importance in a scattering process is suppressed at low enough sound-wave frequencies (lower than the chemical potential). The anti-symmetrized second derivative appearing in the second line does not vanish, due to the non-trivial topology of the vortex solution:
\be
(\di_x \di_y -\di_y \di_x) \psi_V = (2 \pi) \, \delta^2(x,y) \; .
\ee
Precisely the same `topological interaction' term was found in \cite{supersolids} in the context of supersolids.

The interaction Lagrangian further simplifies if we express it directly in terms of the propagating eigenmodes---first and second sound. At leading order in $F_X$ they are given in \eqref{pi1pi2}. We get simply:
\begin{align}
{\cal L}_{\rm int}  & \simeq - F_{yy} y_0 \, \dot \pi_2\,\dot{\vec \pi}_1 \cdot \vec \nabla \psi_V  \nonumber \\
& - \pi \, F_y \, (\dot {\vec \pi}_1 \times \vec \pi_1)\cdot \hat z \,  \delta^2(x,y) \; .
\end{align}
The first line describes the conversion of second sound into first sound, and vice versa. The second line describes the elastic scattering of first sound off the vortex. 
At this order in $F_X$, there is no elastic second sound-vortex scattering---but this  may be deceiving: the second sound speed also vanishes in the $F_X \to 0$ limit. The incoming and outgoing states' propagation speeds enter the scattering cross section in several ways, e.g.~via the phase-space element. As a consequence, it is possible that the second sound-second sound scattering cross section approach a finite limit for $F_X \to 0$. Indeed, \cite{LS} claims that elastic second sound scattering by a superfluid vortex is the dominant process for superfluid helium. Analyzing this possibility is beyond the scope of this paper, and we leave it for future work.

For simplicity, let's restrict to scattering processes where the incoming wave travels orthogonally to the vortex, $\vec k \cdot \hat z =0$. Since the vortex does not break translations along $z$, the $z$-component of the momentum is conserved. This implies that the outgoing wave is also going to propagate perpendicularly to the vortex, $\vec k' \cdot \hat z =0$. This effectively reduces the problem to a two-dimensional one, taking place in the $x$-$y$ plane. Moreover, energy is also conserved, $\omega'= \omega$, because the vortex does not break time translations. 
To compute the associated scattering amplitudes, we apply the standard Feynman rules associated with the so-called relativistic normalization for the single-particle states. As reviewed in \cite{ENRW}, this approach is consistent---and convenient---even when Lorentz-invariance is not there (in our case, it is spontaneously broken by the background). We have to take into account the non-canonical normalization of our first and second sound fields---see eq.~(\ref{L2belowTc}). To simplify the notation, let's denote the combination $(F_y y_0 -F_b b_0)$ by $w$. The relevant (two-dimensional) $S$-matrix elements then are
\begin{align}
i{M}_{1 \to 1} & =  -2 \pi i \, \sfrac{F_y}{w} \cdot \omega \sin \theta \cdot (2\pi) \delta(\omega' - \omega)\\
i{M}_{1 \to 2} & = i{M}_{2 \to 1} \nonumber \\
& \simeq  2 \pi  \, \sqrt \sfrac{F_{yy}}{ {w}} \cdot c_2 \omega \sin \theta \cdot (2\pi) \delta(\omega' - \omega) \; ,
\end{align}
where $\theta$ is the scattering angle.
Since the vortex breaks translations in the $x$-$y$ plane, there is no $\delta$-function enforcing two-dimensional momentum conservation. For the $1\to 2$ and the $2 \to 1$ conversion processes, we have kept the leading order in $c_2/c_1 \ll 1$.

The associated cross-sections per unit vortex length are
\begin{align}
\frac{d \sigma_{a \to b}}{d \theta \, dz} & = \frac{1}{2 \omega} \frac{1}{c_a} \big| {\cal M}_{a \to b} \big|^2 \frac{1}{4\pi c_b^2} \; ,
\end{align}
where $c_{a,b}$ are the initial and final states' propagation speeds, ${\cal M}_{a \to b}$ are the $S$-matrix elements $M_{a \to b}$ without  the $(2\pi) \, \delta(\dots)$ factors, and the last factor comes from the final state's phase space element:
\be
d \Pi_b = \frac{d^2 k'}{(2\pi)^2} \frac{1}{2 \omega'} \cdot (2\pi) \delta(\omega' - \omega) \quad \to \quad \frac{d \theta}{4\pi} \frac{1}{c_b^2} 
\ee
We get
\begin{align}
\frac{d \sigma_{1 \to 1}}{d \theta \, dz} & = \frac{\pi}{2}  \frac{F_{y}^2}{w^2}\cdot \frac{1}{c_1^3} \omega \sin^2 \theta \\
\frac{d \sigma_{1 \to 2}}{d \theta \, dz} & = \frac{\pi}{2}  \frac{F_{yy}}{w}\cdot \frac{1}{c_1} \omega \sin^2 \theta \\
\frac{d \sigma_{2 \to 1}}{d \theta \, dz} & = \frac{c_2}{c_1} \cdot \frac{d \sigma_{1 \to 2}}{d \theta \, dz}
\end{align}
We can express these formulae in terms of thermodynamical quantities. To this end,  recall that for near vanishing $F_X$, we have (see sect.~\ref{thermo})
\be
F_y \simeq n \; , \quad w \equiv F_y y -F_b b \simeq \rho + p \; , \quad F_{yy} \simeq \frac{\di n}{\di \mu} \Big|_s \; .
\ee
We thus get
\begin{align}
\frac{d \sigma_{1 \to 1}}{d \theta \, dz} & \simeq \frac{\pi}{2}  \frac{n^2}{(\rho+p)^2}\cdot \frac{1}{c_1^3} \omega \sin^2 \theta \\
\frac{d \sigma_{1 \to 2}}{d \theta \, dz} & \simeq \frac{\pi}{2}  \frac{(\di n/\di \mu)_s}{(\rho+p)}\cdot \frac{1}{c_1} \omega \sin^2 \theta \\
\frac{d \sigma_{2 \to 1}}{d \theta \, dz} & = \frac{c_2}{c_1} \cdot \frac{d \sigma_{1 \to 2}}{d \theta \, dz}
\end{align}
These results are fully relativistic. In the non-relativistic limit, $\sigma_{1\to1}$ reduces precisely to the elastic phonon-vortex cross-section computed by Son for supersolids \cite{supersolids}.

A final comment is in order: to compute these scattering cross sections we have found it convenient to use a quantum mechanical language.
However our results are entirely classical, and hold for classical sound waves  without any modifications. 
As usual for the scattering of waves off a `potential' or an external source---the vortex, in our case---one can use tree-level quantum perturbation theory to compute the classical cross-section,  defined as the ratio between the scattered power and the incoming energy flux.


\section{Discussion}
We have presented a low-energy, long-distance effective field theory for finite-temperature relativistic superfluids. It is equivalent to more standard hydrodynamical approaches, yet it offers---we believe---a conceptually clearer and technically simpler framework. As emphasized in \cite{DHNS} for ordinary fluids, it may offer a particularly convenient  setup to organize the derivative expansion. Besides, certain properties may be invisible---or at least very well hidden---at the level of hydrodynamical equations, yet manifest at the level of the action, like for instance the unitarity constraints of \cite{DHN}. Moreover,   the Lagrangian formalism is invaluable for  perturbation theory, e.g.~in the systematic study of scattering processes.
As an example we have computed the cross-section for sound wave-vortex scattering. The computation is straightforward, and the  result  easily expressed in terms of hydrodynamical quantities and of the equation of state.
One of the possible applications of our formalism we envisage for the near future, is reproducing the hydrodynamical effects of quantum anomalies \cite{Lin} within effective field theory. This would require supplementing our field theory with Wess-Zumino-like terms, along the lines of \cite{DHN}.

\vspace{.3cm}
 
 \noindent
{\em Acknowledgements.}
I would like to thank Solomon Endlich, Federico Piazza, and Slava Rychkov   for useful discussions and comments. I am especially grateful to Sergei Dubovsky and Lam Hui for enlightening discussions and for collaboration on related subjects. 
This work  is supported by the DOE under contracts DE-FG02-11ER41743 and DE-FG02-92-ER40699, and by NASA under contract NNX10AH14G.


%
%


\end{document}